
%
%
\message{ >>>>> B Paper <<<<< }
\magnification\magstep1
\tracingstats = 1
\message{*** to be used with PLAIN ***}
\newskip\footlineskip \footlineskip = 24pt
\def\makefootline{\baselineskip=\footlineskip \line{\the\footline}}
\def\folio{\ifnum\pageno<-1 \romannumeral-\pageno \fi%
           \ifnum\pageno>0 \number\pageno \fi}
\parskip 1.1ex plus 0.4ex
\parindent 0pt
\pretolerance 1000
\tolerance 2000
\hbadness 2000
\mathsurround 0.2pt
\belowdisplayshortskip = \belowdisplayskip
\advance\hsize -4truemm
%
\def\oddpage{\par\vfill\ifodd\pageno\eject\hbox{\ }\vfill
    \centerline{* * *}\vfill\fi\eject}
%
\def\note#1#2#3{\line{\it #1\hfil \the\day.\the\month.\the\year}
    \vskip 7mm plus 1mm
    {\centerformat{20pt}\bigfonts \bm \bf
    \noindent #2\par}
    \vskip 7mm plus 1mm
    \line{\it\hfil #3\hfil}
    \pageno -1
    \vskip 10mm plus 2mm}
%
\newif\ifnumbers \numberstrue
\def\@{\char"40\relax}
\catcode`@=11 

\font\tensf = cmss10
\font\eightsf = cmss8
\newfam\ssfam\textfont\ssfam=\tensf\scriptfont\ssfam=\eightsf
    \scriptscriptfont\ssfam=\eightsf
\def\sf{\textfont0=\tensf\fam\ssfam\tensf}
\font\mbsym = cmbsy10
\font\mbold = cmmib10
\font\bigrm = cmr10 scaled \magstep2
\font\bigbf = cmbx10 scaled \magstep2
\font\bigit = cmti10 scaled \magstep2
\font\bigtt = cmtt10 scaled \magstep2
\font\bigsl = cmsl10 scaled \magstep2
\font\bigsy = cmsy10 scaled \magstep2
\font\bigbsy = cmbsy10 scaled \magstep2
\font\bigi  = cmmi10 scaled \magstep2
\font\bigmb = cmmib10 scaled \magstep2
\font\bigss = cmss10 scaled \magstep2
\def\bigfonts{\baselineskip 18\p@
    \lineskiplimit 1.8\p@ \lineskip 1.8\p@
    \let\scriptrm\tenrm
    \let\scriptsy\tensy
    \let\scripti\teni
    \let\tenrm\bigrm
    \let\tenbf\bigbf
    \let\tenit\bigit
    \let\tentt\bigtt
    \let\tensl\bigsl
    \let\tensy\bigsy
    \let\mbsym\bigbsy
    \let\mbold\bigmb
    \let\tensf\bigss
    \let\teni\bigi
    \textfont0=\bigrm \textfont1=\bigi \textfont2=\bigsy
    \scriptfont0=\scriptrm \scriptfont1=\scripti \scriptfont2=\scriptsy
    \scriptscriptfont0=\sevenrm
    \scriptscriptfont1=\seveni
    \scriptscriptfont2=\sevensy
    \textfont\itfam=\bigit \textfont\bffam=\bigbf
    \textfont\ttfam=\bigtt \textfont\ssfam=\bigss
    \rm}
\font\ninerm = cmr9
\font\ninebf = cmbx9
\font\nineit = cmti9
\font\ninett = cmtt9
\font\ninesl = cmsl9
\font\ninesy = cmsy9
\font\ninei  = cmmi9
\def\ninefonts{\baselineskip 0.9\baselineskip
    \lineskiplimit 0.9\p@ \lineskip 0.9\p@
    \let\tenrm\ninerm
    \let\tenbf\ninebf
    \let\tenit\nineit
    \let\tentt\ninett
    \let\tensl\ninesl
    \let\tensy\ninesy
    \ifx\ninewasy\relax\else\let\tenwasy\ninewasy\fi
    \let\teni\ninei
    \textfont0=\ninerm \textfont1=\ninei
    \textfont2=\ninesy
    \textfont\itfam=\nineit \textfont\bffam=\ninebf
    \textfont\ttfam=\ninett
    \rm}
%
%
\def\bm{\textfont0=\tenbf\textfont1=\mbold\textfont2=\mbsym
    \ifx\tenwasy\relax\else\let\tenwasy\wasyb\fi}
\def\bmath#1{{\mathsurround \z@\bm\hbox{$#1$}}}

%
%
\catcode`\|=\active \def|{\vert}
\let\3\ss

%
%
\global\lccode`\^^Y=`\^^Y
\gdef\allowhyphens{\penalty\@M \hskip\z@skip}

%
%
\def\frac#1#2{{#1\over#2}}
\def\E{\cdot10^}
%
%
\let\Unitfont\rm
\def\Unit#1{\mskip 2mu{\Unitfont #1}\mskip 1mu}
%
%
\let\?\allowbreak
\def\:{.\kern 2.5\p@\ignorespaces}
\def\0{\phantom{0}}
\def\<#1#2{#1.\kern 1.5\p@#2.\hbox{}} 
\newdimen\t@ntruept \t@ntruept=10truept
\newdimen\@netruept \@netruept=1truept
%
%
\def\ctl#1{\csname #1\endcsname}
\def\ifundefined#1{\expandafter\ifx\csname #1\endcsname\relax}
%
%
\def\vlist#1#2{\par\hangindent#1\hangafter1\noindent{\setbox0
          \hbox{#2\ }\ifdim\wd0<#1\hbox to #1{#2\ \hfil}\else\box0\fi}}
\def\vrlist#1#2{\par\hangindent#1\hangafter1\noindent\hbox to #1{%
          \hss#2\ }}
\def\blist{\vlist{6mm}{$\bullet$}}

\def\newline{\hfill\break}

\def\vdistance#1{\vrule width \z@ #1\relax}
\def\vf{\vfil\penalty 250\vfilneg\penalty \@M}
%
%
\def\boxrulewidth{0.4pt}
\def\boxskip{\vskip\abovedisplayskip}
\long\def\boxtop{\boxskip\vbox\bgroup\offinterlineskip
    \hrule height \boxrulewidth
    \line\bgroup\vrule width \boxrulewidth \hfil\vbox\bgroup
    \advance \hsize by -2em \normalbaselines \parindent \z@
    \vskip 1ex\relax}
\long\def\boxbot{\par\vskip 1ex
    \egroup\hfil \vrule width \boxrulewidth\egroup
    \hrule height \boxrulewidth
    \egroup\boxskip}
%
\def\centerformat#1{%
    \let\newline\break
    \pretolerance = 2000 \tolerance = 3000 \hbadness 500
    \parindent = \z@ \catcode"0D = 5
    \rightskip = #1 plus 1fil \leftskip  = #1 plus 1fil
    \spaceskip = 0.35em \xspaceskip = 0.45em \parfillskip = \z@
    \hyphenpenalty = 9000 \exhyphenpenalty = 5000}
%
%
\def\cbox#1{\setbox0\hbox to \z@{\hss #1\hss}\dp0=\z@\ht0=\z@\box0}
\def\ccbox#1{\vbox to \z@{\vss\hbox to \z@{\hss #1\hss}\vss}}
\def\nbox#1{\hbox to \z@{#1\hss}}
%
%
\def\PgR#1{\expandafter\xdef\csname PgR#1\endcsname%
    {\folio}}
%
%
\def\hide{\begingroup\setbox0\vbox\bgroup}
\def\endhide{\egroup\endgroup}
\catcode`@=12 
%
%
%
\newcount\eqnumber
\newcount\tablecount
\newcount\figcount
\newcount\usernumberi
\newcount\usernumberii
\newcount\usernumberiii
\newbox\toc
\let\hdic\relax
\def\hdrbf{\bm\bf}
\def\hdriibf{\sl}
\def\hdrskip{\vskip 1.3ex}
\def\hdrbreak#1#2#3{%
    \vskip #2 plus #3\goodbreak
    \dimen255=\pagetotal\advance\dimen255 by #3
    \advance\dimen255 by #1\advance\dimen255 by \pagestretch
    \ifdim\dimen255>\pagegoal\vfil\eject\fi}

\def\figname{Fig.\ }
\ifundefined{figrulewidth}\def\figrulewidth{0.4pt}\fi

\def\hdrni#1{%
    \hdrbreak{7truecm}{8ex}{2ex}
    \noindent{\hdrbf #1
    }\par\penalty10000\hdrskip\penalty 10000
    \global\setbox\toc\vbox{\parskip 0pt\rightskip 0pt plus 1fil
    \unvbox\toc\vskip 1.2ex plus 0.8ex
    \noindent\vdistance{height 2ex}{\bm\bf #1}
    \leaders\hbox to 9.5pt{\hfil\rm.\hfil}\hfill
    \vdistance{depth 0.6ex}\folio\par}}

\def\appendix#1{}
\def\hdrii#1{%
    \hdrbreak{10ex}{5ex}{1ex}
    \noindent{\hdriibf #1
    }\par\penalty10000\hdrskip\penalty 10000
    \global\setbox\toc\vbox{\parskip 0pt\rightskip 0pt plus 1fil
    \unvbox\toc
    \noindent\vdistance{height 2ex}#1
    \leaders\hbox to 9.5pt{\hfil\rm.\hfil}\hfill
    \vdistance{depth 0.6ex}\folio\par}}
\def\hdriii#1{%
    \hdrbreak{10ex}{4ex}{1ex}
    \noindent{\hdriibf #1
    }\par\penalty10000\hdrskip\penalty 10000
    \global\setbox\toc\vbox{\parskip 0pt\rightskip 0pt plus 1fil
    \unvbox\toc\hangindent 1em\hangafter 0
    \noindent\vdistance{height 2ex}#1
    \leaders\hbox to 9.5pt{\hfil\rm.\hfil}\hfill
    \vdistance{depth 0.6ex}\folio\par}}
\def\eqnum{\global\advance\eqnumber1
    \eqno({\rm\hdic\the\eqnumber})}
\def\theeqnum{\global\advance\eqnumber1{\rm\hdic\the\eqnumber}}
\def\QnR#1{\expandafter\xdef\csname QnR#1\endcsname%
    {{\rm\hdic\the\eqnumber}}}
\def\figtop#1{\vbox\bgroup\offinterlineskip
    \hrule height \figrulewidth
    \line{\vrule width \figrulewidth height #1\hfil\vrule%
    width \figrulewidth}}
\def\figspace#1{\line{\vrule width \figrulewidth height #1\hfil\vrule%
    width \figrulewidth}}
\long\def\figbody#1{\line{\vrule width \figrulewidth\relax #1\hss\vrule%
    width \figrulewidth}}
\long\def\startfigbody{\vskip 0pt
     \line\bgroup\vrule width \figrulewidth\relax}
\long\def\endfigbody{\hfil\vrule width \figrulewidth\egroup}
\def\figcap#1{\line{\vrule width \figrulewidth
    \ninefonts
    \hfil\ifnumbers\fignum\hfil\fi
    \vtop{\ifnumbers\advance\hsize by -7.5em\else
    \advance\hsize by -1.5em\fi
    \normalbaselines\noindent#1\vphantom{p)}\par
    \kern 3.6pt}\hfil\vrule width \figrulewidth}}
\def\fignum{\global\advance\figcount 1
    {\bf \figname \hdic\the\figcount}}
\def\figxcap#1#2{%
    \line{\vrule width \figrulewidth \hfil
    {\bf #1}\hfil
    \vtop{\advance\hsize by -7.5em
    \normalbaselines\noindent#2\vphantom{p)}\par
    \kern 3.6pt}\hfil\vrule width \figrulewidth}}
\def\figbot{\hrule height \figrulewidth \egroup}
\def\FiG#1{\expandafter\xdef\csname FiG#1\endcsname%
    {\hdic\the\figcount}}
\def\nextFiG{{\count0=\figcount\advance\count0 by1\relax
    \hdic\the\count0}}
%
%
%
\catcode`@=11
\font\circlefont = lcircle10 at \t@ntruept
\font\linefont = line10 at \t@ntruept
\def\put(#1,#2)#3{\raise #2 truept\hbox to \z@%
    {\kern #1\@netruept\relax#3\hss}}
\def\putl(#1,#2)#3{\raise #2\@netruept\hbox to \z@%
    {\kern #1\@netruept\llap{#3}\hss}}
\def\circUR#1{{\count1=#1\relax\count0=#1\multiply\count0 by 2
    \hbox{%
    \advance\count0by-4\circlefont\char\count0}}}
\def\circUL#1{{\count1=#1\relax\count0=#1\multiply\count0 by 2
    \hbox{%
    \advance\count0by-1\circlefont\char\count0}}}
\def\circLR#1{{\count1=#1\relax\count0=#1\multiply\count0 by 2
    \hbox{%
    \advance\count0by-3\circlefont\char\count0}}}
\def\circLL#1{{\count1=#1\relax\count0=#1\multiply\count0 by 2
    \hbox{%
    \advance\count0by-2\circlefont\char\count0}}}
\def\circle#1{{\count0=#1\advance\count0\count0
    \raise\count0truept \hbox to \z@{%
    \circUL{#1}\circUR{#1}\hss}\circLL{#1}%
    \circLR{#1}}}
\def\arrR{{\linefont\char"2D}}
\def\arrL{{\linefont\char"1B}}
\def\arrU{{\linefont\char"36}}
\def\arrD{{\linefont\char"3F}}
\newbox\feyndashbox
\newbox\feyncircubox
\newbox\feyncirclbox
\newbox\feyngluebox
\setbox\feyndashbox\hbox{%
    \kern 4\@netruept\vrule width 4\@netruept height 0.2\@netruept
    depth 0.2\@netruept}
\setbox\feyncircubox\hbox{%
    \raise 2\@netruept \hbox to 4\@netruept{\circUL{2}\circUR{2}\hss}}
\setbox\feyncirclbox\hbox{%
    \lower 2\@netruept \hbox to 4\@netruept{\circLL{2}\circLR{2}\hss}}
\setbox\feyngluebox\hbox{%
    \hbox to \z@{\circLL{2}\hss}\hbox to 2\@netruept{\circUL{2}\hss}%
    \vrule height 0.2\@netruept depth  0.2\@netruept width 3\@netruept}
\def\feynloop{\raise 2\t@ntruept \hbox to \z@{\circUL{20}%
    \circUR{20}\hss}\lower 2\t@ntruept%
    \hbox to 4\t@ntruept{\circLL{20}\circLR{20}\hss}}
\def\feynlloop{\raise 2\t@ntruept \hbox to \z@{\circUL{20}\hss}%
    \lower 2\t@ntruept \hbox to 2\t@ntruept{\circLL{20}\hss}}
\def\feynrloop{\raise 2\t@ntruept \hbox to \z@{%
    \kern 2\t@ntruept\circUR{20}\hss}%
    \lower 2\t@ntruept \hbox to 2\t@ntruept{%
    \kern 2\t@ntruept\circLR{20}\hss}}
\def\feynline{\vrule width 4\t@ntruept height 0.2\@netruept
    depth 0.2\@netruept}
\def\feynlineR{\hbox to \z@{\kern 1.5\t@ntruept\arrR\hss}%
    \vrule width 4\t@ntruept height 0.2\@netruept depth 0.2\@netruept}
\def\feynlineL{\hbox to \z@{\kern 1.5\t@ntruept\arrL\hss}%
    \vrule width 4\t@ntruept height 0.2truept depth 0.2truept}
\def\feynxline#1{\vrule width #1\@netruept height 0.2\@netruept
    depth 0.2\@netruept}
\def\feynvline{\hbox to \z@{%
    \hss\vrule width 0.4\@netruept height 4\t@ntruept depth \z@\hss}}
\def\feynvlineU{\hbox to \z@{\raise 1.5\t@ntruept \hbox to \z@{%
    \arrU\hss}%
    \hss\vrule width 0.4\@netruept height 4\t@ntruept depth \z@\hss}}
\def\feynvlineD{\hbox to \z@{\raise 1.5\t@ntruept \hbox to \z@{%
    \arrD\hss}%
    \hss\vrule width 0.4\@netruept height 4\t@ntruept depth \z@\hss}}
\def\feynascend{\hbox{\linefont\char0\raise \t@ntruept \hbox{%
    \char0}\raise 2\t@ntruept\hbox{\char0}\raise 3\t@ntruept\hbox{%
    \char0}}}
\def\feynascendU{\hbox{\linefont\char0\raise \t@ntruept \hbox{%
    \hbox to \z@{\char"12\hss}%
    \char0}\raise 2\t@ntruept\hbox{\char0}\raise 3\t@ntruept\hbox{%
    \char0}}}
\def\feynascendD{\hbox{\linefont\char0\raise \t@ntruept \hbox{%
    \char0}\raise 2\t@ntruept\hbox{%
    \hbox to \z@{\char"09\hss}%
    \char0}\raise 3\t@ntruept\hbox{\char0}}}
\def\feyndescendU{\hbox{\linefont\lower \t@ntruept \hbox{%
    \char64}\lower 2\t@ntruept\hbox{\char64}\lower 3\t@ntruept\hbox{%
    \hbox to \z@{\char"49\hss}%
    \char64}\lower 4\t@ntruept\hbox{\char64}}}
\def\feyndescendD{\hbox{\linefont\lower \t@ntruept \hbox{%
    \char64}\lower 2\t@ntruept\hbox{\hbox to \z@{\char"52\hss}%
    \char64}\lower 3\t@ntruept\hbox{%
    \char64}\lower 4\t@ntruept\hbox{\char64}}}
\def\feyndescend{\hbox{\linefont\lower \t@ntruept \hbox{%
    \char64}\lower 2\t@ntruept\hbox{\char64}\lower 3\t@ntruept\hbox{%
    \char64}\lower 4\t@ntruept\hbox{\char64}}}
\def\feyncross{\hbox{\linefont\raise 3\t@ntruept \hbox to \z@{%
    \char64\hss}\char0\raise 2\t@ntruept\hbox{%
    \char64}\raise \t@ntruept\hbox{%
    \char64}\raise 3\t@ntruept\hbox to \z@{\char0\hss}\char64}}
\def\feyndash{\vrule width 2truept height 0.2truept depth 0.2truept
    \copy\feyndashbox%
    \copy\feyndashbox%
    \copy\feyndashbox%
    \copy\feyndashbox%
    \kern 4\@netruept\vrule width 2truept height 0.2truept depth 0.2truept}
\def\feynvdash{\hbox to \z@{%
    \hss\vrule width 0.4\@netruept height 2\@netruept depth \z@\hss}%
    \hbox to \z@{\hss
    \vrule width 0.4\@netruept height\t@ntruept depth -6\@netruept\hss}%
    \hbox to \z@{\hss
    \vrule width 0.4\@netruept height 18\@netruept
    depth -14\@netruept\hss}%
    \hbox to \z@{%
    \hss\vrule width 0.4\@netruept height 26\@netruept
    depth -22\@netruept\hss}%
    \hbox to \z@{%
    \hss\vrule width 0.4\@netruept height 34\@netruept
    depth -3\t@ntruept\hss}%
    \hbox to \z@{%
    \hss\vrule width 0.4\@netruept height 4\t@ntruept
    depth -38\@netruept\hss}}
\def\feynphot{\setbox0\hbox{%
    \copy\feyncircubox\copy\feyncirclbox%
    \copy\feyncircubox\copy\feyncirclbox%
    \copy\feyncircubox\copy\feyncirclbox%
    \copy\feyncircubox\copy\feyncirclbox%
    \copy\feyncircubox\copy\feyncirclbox}\ht0=\z@\dp0=\z@\box0}
\def\feynvphot{\setbox0\hbox to \z@{\hss\kern 4\@netruept
    \vbox to 4\t@ntruept{\offinterlineskip\vss
    \circUL{2}
    \hbox{\circLL{2}\circUR{2}}\hbox{\circUL{2}\circLR{2}}
    \hbox{\circLL{2}\circUR{2}}\hbox{\circUL{2}\circLR{2}}
    \hbox{\circLL{2}\circUR{2}}\hbox{\circUL{2}\circLR{2}}
    \hbox{\circLL{2}\circUR{2}}\hbox{\circUL{2}\circLR{2}}
    \hbox{\circLL{2}\circUR{2}}\hbox{\kern 4\@netruept\circLR{2}}
    }\hss}\dp0=\z@\box0}
\def\feynglue{\setbox0\hbox{%
    \copy\feyncircubox\hbox to \z@{\hss%
    \vrule height \z@ depth 3\@netruept width 0.4\@netruept\hss}%
    \copy\feyncircubox\hbox to \z@{\hss%
    \vrule height \z@ depth 3\@netruept width 0.4\@netruept\hss}%
    \copy\feyncircubox\hbox to \z@{\hss%
    \vrule height \z@ depth 3\@netruept width 0.4\@netruept\hss}%
    \copy\feyncircubox\hbox to \z@{\hss%
    \vrule height \z@ depth 3\@netruept width 0.4\@netruept\hss}%
    \copy\feyncircubox\hbox to \z@{\hss%
    \vrule height \z@ depth 3\@netruept width 0.4\@netruept\hss}%
    \copy\feyncircubox\hbox to \z@{\hss%
    \vrule height \z@ depth 3\@netruept width 0.4\@netruept\hss}%
    \copy\feyncircubox\hbox to \z@{\hss%
    \vrule height \z@ depth 3\@netruept width 0.4\@netruept\hss}%
    \copy\feyncircubox\hbox to \z@{\hss%
    \vrule height \z@ depth 3\@netruept width 0.4\@netruept\hss}%
    \copy\feyncircubox\hbox to \z@{\hss%
    \vrule height \z@ depth 3\@netruept width 0.4\@netruept\hss}%
    \copy\feyncircubox}\ht0=\z@\dp0=\z@\box0}
\def\feynvglue{\setbox0\hbox to \z@{\hss\kern 4\@netruept
    \vbox to 4\t@ntruept{\offinterlineskip\vss
    \circUL{2}
    \copy\feyngluebox
    \copy\feyngluebox
    \copy\feyngluebox
    \copy\feyngluebox
    \copy\feyngluebox
    \copy\feyngluebox
    \copy\feyngluebox
    \copy\feyngluebox
    \copy\feyngluebox
    \circLL{2}
    }\hss}\dp0=\z@\box0}
\def\feynlv{\raise 4\t@ntruept \hbox to \z@{\feyndescend\hss}%
    \lower 4\t@ntruept \hbox {\feynascend}}
\def\feynrv{\hbox to \z@{\feyndescend\hss}\feynascend}
\catcode`@=12 
\newtoks\tabpreamble
\newskip\tabcenter \tabcenter=0pt plus 40pt
\newif\iftabnumbers \ifnumbers\tabnumberstrue\else\tabnumbersfalse\fi
\def\tabrulewidth{0.4pt}
\def\tabbotruleheight{0.8pt}
\def\hdic{}
\let\tabhbox\line
\def\tabname{Table\ }
\ifundefined{fnfont}\gdef\fnfont{\ninefonts}\fi
\ifundefined{tabfont}\gdef\tabfont{}\fi
\def\tabspace{}
\def\tabfinrule{\tabspace\noalign{\nobreak\hrule\nobreak}}
\def\tabrule{\tabfinrule\tabspace}

\def\tabtopskip{\vskip 10pt plus 4pt minus 7pt}
\long\def\tabstart#1{\tabtopskip
    \bgroup\offinterlineskip
    \everycr{\noalign{\nobreak}}
    \def\1{\span\omit}
    \halign to\hsize \bgroup \vrule ##&
    ##\tabskip \tabcenter&\strut #1 ##\tabskip 0pt\vrule\cr
    \noalign{\nobreak\hrule height \tabrulewidth\nobreak}}
\def\hash{####}
\long\def\tabtop#1{%
    {\def|{\tabDEF1}
    \def\tabvrule##1{\tabDEF1}
    \def\?{\tabDEF2}
    \let\tabDEF0
    \xdef\next{\tabDEF0#1\tabDEF3}}
    \tabpreamble={}
    \def\tabDEF##1##2\tabDEF##3{%
        \ifx##11\tabpreamble=\expandafter{\the\tabpreamble&}\fi
        \ifx##12\tabpreamble=\expandafter{\the\tabpreamble&\omit}\fi
        \ifx##33\def\next{}\else\def\next{\tabDEF##3}\fi\next}
    \next
    \edef\tabspace{height 2.0pt&\the\tabpreamble&\cr}
    {\def|{\vrule\hash&}
    \def\tabvrule##1{\vrule width##1\hash&}
    \def\?{\hash}
    \xdef\next{\tabpreamble={#1}}}\next
    \expandafter\tabstart\expandafter{\the\tabpreamble}
    }
\def\tabcap#1{\noalign{\global\advance\tablecount 1}\tabcaption{#1}}

\def\tabcont#1{\noalign{%
    \tabhbox{\vrule width \tabrulewidth height 2.4pt\hfil
    \vrule width \tabrulewidth}
    \nobreak
    \tabhbox{\vrule width \tabrulewidth\tabfont\hfil\iftabnumbers
    \phantom{\bf
    \tabname\hdic\the\tablecount}\hfil\fi
    \vtop{\iftabnumbers\advance\hsize by -8em
    \else\advance\hsize by -1em\fi
    \normalbaselines\tabfont
    \noindent#1\vphantom{p}\par\kern 3.6pt}\hfil
    \vrule width \tabrulewidth}
    \nobreak\hrule\nobreak}}
\def\tabcaption#1{\noalign{%
    \tabhbox{\vrule width \tabrulewidth height 2.4pt\hfil
    \vrule width \tabrulewidth}
    \nobreak
    \tabhbox{\vrule width \tabrulewidth\tabfont\hfil\iftabnumbers{\bf
    \tabname\hdic\the\tablecount}\hfil\fi
    \vtop{\iftabnumbers\advance\hsize by -8em
    \else\advance\hsize by -1em\fi
    \normalbaselines\tabfont
    \noindent#1\vphantom{p}\par\kern 3.6pt}\hfil
    \vrule width \tabrulewidth}
    \nobreak\hrule\nobreak}}

\def\tabfn#1#2{\noalign{\nobreak\tabhbox{\vrule width \tabrulewidth
    \vdistance{height 10pt}\hfil
    \vtop{\advance\hsize by -2em
    \normalbaselines\fnfont\vrlist{1.2em}{$^{#1}$}%
    #2\vphantom{p}\par}\hfil\vrule width \tabrulewidth}
    \nobreak
    \tabhbox{\vrule  width \tabrulewidth height 2.0pt\hfil
    \vrule width \tabrulewidth}}}
\def\tabbot{\noalign{\nobreak\hrule height \tabbotruleheight}%
    \egroup\egroup}

\iftabnumbers
\def\TaB#1{\expandafter\xdef\csname TaB#1\endcsname%
    {\hdic\the\tablecount}}
\def\nextTaB{{\count0=\tablecount\advance\count0 by1\relax
    \hdic\the\count0}}
\else
\def\TaB#1{\expandafter\xdef\csname TaB#1\endcsname%
    {above}}
\def\nextTaB{below}
\fi
\newcount\citnumber
\def\ReF#1{\def\rEf{#1}\ifx\rEf\empty\else%
           \expandafter\ifx\csname ReF#1\endcsname\relax%
           \global\advance\citnumber 1%
           \expandafter\xdef\csname ReF#1\endcsname%
           {\the\citnumber}\fi\csname ReF#1\endcsname\fi}
\def\cit#1{{\tt\char"5B}#1{\tt\char"5D}}         
\def\cite#1{\cit{\ReF{#1}}}
\def\CiT#1#2,#3(#4){{\frenchspacing #1 {\bf\ignorespaces #2}%
, {\ignorespaces #3\unskip} (#4)}} 
\def\PRL{\CiT{Phys. Rev. Lett.}}
\def\PL{\CiT{Phys. Lett.}}

\def\PR{\CiT{Phys. Rev.}}
\def\ZP{\CiT{Z. Phys.}}

\def\bibitem#1{\vlist{1cm}{{\tt\char"5B}\ReF{#1}{\tt\char"5D}}
    \ignorespaces}                                            
%
%

\def\BR{\mathop{\rm BR}}

\def\Y#1S{{\Upsilon\rm(#1S)}}
\def\subdecay{\hbox{\vrule height 7pt depth -2pt
    \kern -1pt\tensy\char"21}}

\def\M#1#2#3{\ifmmode{}^{#1}{\rm #2}_{#3}\else
    \hbox{${}^{#1}{\rm #2}_{#3}$}\fi}

\def\alphas{\alpha_{\rm s}}
\def\Jpsi{{J\mskip -3mu/\mskip -2mu\psi\mskip 2mu}}

\def\parall{{\mathchoice{}{}
    {\hbox{\kern 0.2ex\vrule height 1.1ex depth 0pt
    \vdistance{height 1.3ex depth 0.1ex}%
    \kern 0.21ex\vrule height 1.1ex depth 0pt\kern 0.2ex}}
    {\hbox{\kern 0.18ex\vrule height 0.8ex depth 0pt
    \vdistance{height 0.90ex depth 0.1ex}%
    \kern 0.18ex\vrule height 0.8ex depth 0pt\kern 0.18ex}}}}

\mathchardef\Gamma="7100
\def\Gam{{\rm\Gamma}}

\mathchardef\Delta="7101
\mathchardef\Theta="7102
\mathchardef\Lambda="7103
\mathchardef\Xi="7104
\mathchardef\Pi="7105
\mathchardef\Sigma="7106
\mathchardef\Upsilon="7107
\mathchardef\Phi="7108
\mathchardef\Psi="7109
\mathchardef\Omega="710A

\def\dbar{\bar d}
\def\sbar{\bar s}
\def\cbar{\bar c}
\def\bbar{\bar b}

\def\Kbar{{\kern 0.2em\overline{\kern -0.2em K}}{}}
\def\Bbar{{\kern 0.18em\overline{\kern -0.18em B}}{}}
\def\Dbar{{\kern 0.2em\overline{\kern -0.2em D}}{}}
\def\anu{{\kern 0.06em\overline{\kern -0.06em \nu}}{}}
\def\pbar{{\kern 0.06em\overline{\kern -0.06em p}}{}}
\def\nbar{{\kern 0.06em\overline{\kern -0.06em n}}{}}
\def\Nbar{{\kern 0.2em\overline{\kern -0.2em N}}{}}
\def\Deltabar{{\kern 0.25em\overline{\kern -0.25em \Delta}}{}}
\def\Lbar{{\kern0.2em\overline{\kern-0.2em\Lambda\kern0.05em}%
    \kern-0.05em}{}}
\def\Sbar{{\kern 0.2em\overline{\kern -0.2em \Sigma}}{}}
\def\Xibar{{\kern 0.2em\overline{\kern -0.2em \Xi}}{}}
\def\Obar{{\kern 0.2em\overline{\kern -0.2em \Omega}}{}}

\def\se{\Unit{s}}

\def\eV{\Unit{e\kern -0.10em V}}
\def\keV{\Unit{ke\kern -0.08em V}}
\def\MeV{\Unit{Me\kern -0.08em V}}
\def\GeV{\Unit{Ge\kern -0.08em V}}
\def\TeV{\Unit{Te\kern -0.08em V}}
\def\byc{\mkern -2mu{/}c}

\def\LQCD{\Lambda_{\rm QCD}}
\def\Nc{N_{\rm c}}
\def\BR{{\cal B}}
\def\tabrulewidth{0pt}
\def\tabfont{\ninefonts}

\catcode`\@=11
\font\linefont = line10 at 10truept  
\newif\ifhistopolyline
\newdimen\histowidth
\newdimen\histoheight
\dimendef\dimen@@=3
\dimendef\dimen@@i=4
\newcount\histoxmin
\newcount\histoxdif
\newcount\histoymin
\newcount\histoydif
\newcount\count@@
\newdimen\histox
\newdimen\histoy
\newdimen\histoxx
\newdimen\histoyy
\newdimen\histoxu
\newdimen\histoyu
\newdimen\histobarwidth
\newdimen\histotickwidth
\newdimen\historulewidth
%
%
\newdimen\plotpointradius
\newdimen\plotdotspacing
\historulewidth=0.8\p@
\histobarwidth=0.4\p@
\histotickwidth=0.4\p@
\def\plotsymbol{\bullet}
\def\histogram#1#2#3#4#5#6{\histopolylinefalse\histowidth=#1\relax
    \histoheight=#2\relax\lower\historulewidth\hbox to \historulewidth{%
    \vbox{\offinterlineskip
    \hsize\histowidth
    \advance\hsize\historulewidth
    \advance\hsize\historulewidth
    \hrule height \historulewidth
    \hbox to \hsize{\vrule height \histoheight width \historulewidth
     \hfil\vrule width \historulewidth}
    \hrule height \historulewidth}\hss}%
    \histoxmin=#3\relax
    \histoxdif=#4\advance\histoxdif by -\histoxmin
    \histox=\histowidth
    \divide\histox by \histoxdif
    \histoymin=#5\relax
    \histoydif=#6\advance\histoydif by -\histoymin
    \histoy=\histoheight
    \divide\histoy by \histoydif
    \histoxu=\histox \histoyu=\histoy
    \ignorespaces}
\def\histocross(#1+#2-#3,#4+#5-#6){%
    \hbox to \z@{%
    {\histoxx=\histox \dimen@i=\histox
     \count@=#1\advance\count@ by -\histoxmin
     \advance\count@ by -#3
     \multiply\histox by \count@
     \kern\histox
     \count@=#4\advance\count@ by -\histoymin
     \multiply\histoy by \count@
     \advance\histoy by -0.5\histobarwidth
     \dimen@=\histoy \advance\dimen@ by \histobarwidth
     \count@@=#3 \advance\count@@ by #2
     \multiply\histoxx by \count@@
     \ifhistopolyline\advance\histoxx by 0.25\histobarwidth\fi
     \vrule height \dimen@ depth -\histoy width \histoxx
     \multiply\dimen@i by #2
     \ifhistopolyline\advance\dimen@i by 0.75\histobarwidth
     \else
     \advance\dimen@i by 0.5\histobarwidth
     \fi
     \kern -\dimen@i}%
    {\histoyy=\histoy
     \count@=#4\advance\count@ by -\histoymin
     \count@@=\count@
     \advance\count@ by -#6
     \advance\count@@ by #5
     \multiply\histoyy by \count@@
     \multiply\histoy by \count@
     \ifhistopolyline\advance\histoyy by 0.25\histobarwidth\fi
     \vrule height \histoyy depth -\histoy width \histobarwidth}%
    \hss}\ignorespaces}
\def\histohline(#1+#2-#3,#4){%
    \hbox to \z@{%
    {\histoxx=\histox \dimen@i=\histox
     \count@=#1\advance\count@ by -\histoxmin
     \advance\count@ by -#3
     \multiply\histox by \count@
     \kern\histox
     \count@=#4\advance\count@ by -\histoymin
     \multiply\histoy by \count@
     \advance\histoy by -0.5\histobarwidth
     \dimen@=\histoy \advance\dimen@ by \histobarwidth
     \count@@=#3 \advance\count@@ by #2
     \multiply\histoxx by \count@@
     \vrule height \dimen@ depth -\histoy width \histoxx
     \multiply\dimen@i by #2
     \kern -\dimen@i \kern -0.5\histobarwidth}%
    \hss}\ignorespaces}
\def\histovline(#1,#2+#3-#4){%
    \hbox to \z@{%
    {\count@=#1\advance\count@ by -\histoxmin
     \multiply\histox by \count@
     \advance\histox by -0.5\histobarwidth
     \kern\histox
     \histoyy=\histoy
     \count@=#2\advance\count@ by -\histoymin
     \count@@=\count@
     \advance\count@ by -#4
     \advance\count@@ by #3
     \multiply\histoyy by \count@@
     \multiply\histoy by \count@
     \vrule height \histoyy depth -\histoy width \histobarwidth}%
    \hss}\ignorespaces}
\def\histopoint(#1,#2){%
    \hbox to \z@{%
    {\count@=#1\advance\count@ by -\histoxmin
     \multiply\histox by \count@
     \kern\histox
     \count@=#2\advance\count@ by -\histoymin
     \multiply\histoy by \count@
     \advance\histoy by -2.5\p@
     \raise\histoy\hbox to \z@{\hss$\plotsymbol$\hss}\hss}}%
     \ignorespaces}
\def\histoput(#1,#2)#3{%
    \hbox to \z@{%
    {\count@=#1\advance\count@ by -\histoxmin
     \multiply\histox by \count@
     \kern\histox
     \count@=#2\advance\count@ by -\histoymin
     \multiply\histoy by \count@
     \raise\histoy\hbox to \z@{#3\hss}\hss}}%
     \ignorespaces}
\def\histoxtick#1#2#3{%
    {\count@=#1\advance\count@ by -\histoxmin
     \multiply\histox by \count@
     \advance\histox by -0.5\histotickwidth
     \hbox to \z@{\kern\histox
     \vrule height #2 width \histotickwidth \hss}%
     \dimen@=\histoheight \advance \dimen@ by -#2
     \hbox to \z@{\kern\histox
     \vrule height \histoheight depth -\dimen@ width \histotickwidth
     \lower 2.3ex\hbox to \z@{\hss\ignorespaces#3\hss}%
     \hss}%
    }\ignorespaces}
\def\histoytick#1#2#3{%
    {\count@=#1\advance\count@ by -\histoymin
     \multiply\histoy by \count@
     \advance\histoy by -0.5\histotickwidth
     \dimen@=\histoy \advance\dimen@ by \histotickwidth
     \hbox to \z@{%
     \vrule height \dimen@ depth -\histoy width #2\hss}%
     \hbox to \z@{\kern\histowidth \kern-#2
     \vrule height \dimen@ depth -\histoy width #2\hss}%
     \advance\histoy by -0.5ex
     \raise\histoy\hbox to \z@{\hss #3\kern 0.4em}%
    }\ignorespaces}
%
%
\def\plotthick{%
    \plotpointradius=38\p@ \divide\plotpointradius by 72
    \plotdotspacing=1.4\plotpointradius
    \gdef\plotpointbox{%
    \lower \plotpointradius\hbox to \z@{\hss\tenrm.\hss}}%
    \ignorespaces}
\def\plotnormal{%
    \plotpointradius=29\p@ \divide\plotpointradius by 72
    \plotdotspacing=1.4\plotpointradius
    \gdef\plotpointbox{%
    \lower \plotpointradius\hbox to \z@{\hss\sevenrm.\hss}}%
    \ignorespaces}
\def\plotthin{%
    \plotpointradius=22\p@ \divide\plotpointradius by 72
    \plotdotspacing=1.4\plotpointradius
    \gdef\plotpointbox{%
    \lower \plotpointradius\hbox to \z@{\hss\fiverm.\hss}}%
    \ignorespaces}
\def\plotline#1#2#3#4{%
    {\dimen@i=#1\dimen@ii=#2\dimen@@=#3\dimen@@i=#4\relax
    \advance\dimen@@ by -\dimen@i\advance\dimen@@i by -\dimen@ii
    \ifdim\dimen@@<\z@ \dimen5=-\dimen@@ \else \dimen5=\dimen@@ \fi
    \ifdim\dimen@@i<\z@ \dimen@=-\dimen@@i \else \dimen@=\dimen@@i \fi{}%
    \ifdim\dimen@>\dimen5 \dimen5=\dimen@ \fi{}%
    \ifdim\dimen5>\plotdotspacing
      \divide\dimen5 by \plotdotspacing  
      \count@=\dimen5 \advance\count@ by 1
      \divide\dimen@@ by \count@ \divide \dimen@@i by \count@
      \loop
      \raise\dimen@ii\hbox to \z@{\kern\dimen@i\plotpointbox\hss}%
      \advance\dimen@i by \dimen@@
      \advance\dimen@ii by \dimen@@i
      \advance\count@ by -1
      \ifnum\count@ > 0 \repeat
    \else
      \advance\dimen@i by 0.5\dimen@@
      \advance\dimen@ii by 0.5\dimen@@i
      \raise\dimen@ii\hbox to \z@{\kern\dimen@i\plotpointbox\hss}%
    \fi}\ignorespaces}
\def\histoline(#1,#2)(#3,#4){%
    \global\histoxx=\histox \global\histoyy=\histoy
    {\count@=#1\advance\count@ by -\histoxmin
     \multiply\histox by \count@
     \count@=#3\advance\count@ by -\histoxmin
     \global\multiply\histoxx by \count@
     \count@=#2\advance\count@ by -\histoymin
     \multiply\histoy by \count@
     \count@=#4\advance\count@ by -\histoymin
     \global\multiply\histoyy by \count@
     \plotline{\histox}{\histoy}{\histoxx}{\histoyy}}\ignorespaces}
\def\histomove(#1,#2)#3\histodraw#4({\histoline(#1,#2)(}
\def\histodraw(#1,#2){%
    {\count@=#1\advance\count@ by -\histoxmin
     \multiply\histox by \count@
     \count@=#2\advance\count@ by -\histoymin
     \multiply\histoy by \count@
     \plotline{\histoxx}{\histoyy}{\histox}{\histoy}
     \global\histoxx=\histox \global\histoyy=\histoy
    }\ignorespaces}
\def\histoascend(#1,#2)(#3){%
    \global\histoxx=\histox \global\histoyy=\histoy
    {\count@=#1\advance\count@ by -\histoxmin
     \multiply\histox by \count@
     \count@=#3\advance\count@ by -\histoxmin
     \global\multiply\histoxx by \count@
     \count@=#2\advance\count@ by -\histoymin
     \multiply\histoy by \count@
     \plotascend{\histox}{\histoy}{\histoxx}}\ignorespaces}
\def\plotascend#1#2#3{%
    {\dimen@i=#1\dimen@ii=#2\dimen@@=#3\relax
    \advance\dimen@@ by -\dimen@i
    \ifdim\dimen@@<\z@ \dimen5=-\dimen@@ \else \dimen5=\dimen@@ \fi
    \dimen@@i = \t@ntruept
    \ifdim\dimen5>\dimen@@i
      \divide\dimen5 by \dimen@@i   
      \count@=\dimen5 \count@@=\count@ \advance\count@ by 1
      \advance\dimen@@ by -\count@\dimen@@i \divide\dimen@@ by \count@@
      \advance\dimen@@ by \dimen@@i
      \loop
      \raise\dimen@ii\hbox to \z@{\kern\dimen@i\linefont\char0\hss}%
      \advance\dimen@i by \dimen@@
      \advance\dimen@ii by \dimen@@
      \advance\count@ by -1
      \ifnum\count@ > 0 \repeat
    \fi}\ignorespaces}
\def\histodescend(#1,#2)(#3){%
    \global\histoxx=\histox \global\histoyy=\histoy
    {\count@=#1\advance\count@ by -\histoxmin
     \multiply\histox by \count@
     \count@=#3\advance\count@ by -\histoxmin
     \global\multiply\histoxx by \count@
     \count@=#2\advance\count@ by -\histoymin
     \multiply\histoy by \count@
     \plotdescend{\histox}{\histoy}{\histoxx}}\ignorespaces}
\def\plotdescend#1#2#3{%
    {\dimen@i=#1\dimen@ii=#2\dimen@@=#3\relax
    \advance\dimen@@ by -\dimen@i
    \ifdim\dimen@@<\z@ \dimen5=-\dimen@@ \else \dimen5=\dimen@@ \fi
    \dimen@@i = \t@ntruept
    \ifdim\dimen5>\dimen@@i
      \divide\dimen5 by \dimen@@i   
      \count@=\dimen5 \count@@=\count@ \advance\count@ by 1
      \advance\dimen@@ by -\count@\dimen@@i \divide\dimen@@ by \count@@
      \advance\dimen@@ by \dimen@@i
      \advance\dimen@ii by -\dimen@@i
      \loop
      \raise\dimen@ii\hbox to \z@{\kern\dimen@i\linefont\char64\hss}%
      \advance\dimen@i by \dimen@@
      \advance\dimen@ii by -\dimen@@
      \advance\count@ by -1
      \ifnum\count@ > 0 \repeat
    \fi}\ignorespaces}
\def\plotsample#1{\hbox{#1%
    \histowidth=6.9truemm \advance \histowidth by \plotdotspacing
    \plotline{0mm}{0.6ex}{\histowidth}{0.6ex}\kern 7truemm}}
\catcode`\@=12 
\plotnormal
\ifundefined{figrulewidth}\def\figrulewidth{0pt}\fi

{
\parskip = 1.1truemm plus 0.25fil
\nopagenumbers
\def\centerformat{%
   \pretolerance = 2000 \tolerance = 3000 \hbadness 500
   \parindent = 0pt \catcode"0D = 5
   \rightskip = 20truept plus 1fil \leftskip  = 20truept plus 1fil
   \spaceskip = 0.35em \xspaceskip = 0.45em \parfillskip = 0pt
   \hyphenpenalty = 1000 \exhyphenpenalty = 1000}
\def\:{.\kern 0.2em\relax}
\font\bigbf = cmbx10 at 14.4truept
\font\bigmb = cmmib10 at 14.4truept
\font\mbold = cmmib10
\font\bigbsy = cmbsy10 at 14.4truept
\font\mbsym = cmbsy10
 at 10.95truept
 at 10.95truept
\font\eightrm  = cmr8 at 8truept
\let\authfont\rm 
\let\instfont\it 
\def\title#1{{\bigbf\centerformat{%
   \textfont0=\bigbf \textfont1=\bigmb \textfont2=\bigbsy
   \scriptfont0=\tenbf \scriptfont1=\mbold \scriptfont2=\mbsym
   \scriptscriptfont0=\sevenrm
   \scriptscriptfont1=\seveni
   \scriptscriptfont2=\sevensy
   \baselineskip = 1.4\normalbaselineskip
   \lineskip = 1.4\normallineskip
   \bigbf #1 \par}}%
   \vskip 2.0truemm plus 0.50fil}
\def\authors#1{{\authfont\centerformat{ #1 \par}}}
\def\inst#1{{\instfont\centerformat\parskip 0pt{ #1 \par}}}
\newcount\fnnumber
\def\fn#1#2{\/%
   \expandafter\ifx\csname FnN#1\endcsname\relax%
   \global\advance\fnnumber 1%
   \expandafter\xdef\csname FnN#1\endcsname%
   {\the\fnnumber}%
   \insert\footins{\interlinepenalty=\interfootnotelinepenalty
   \leftskip=0pt\rightskip=0pt\parfillskip=0pt plus 1fill
   \spaceskip=2.5pt plus 1.5pt minus 1pt
   \xspaceskip=3pt plus 2pt minus 1pt
   \baselineskip = 0.8\normalbaselineskip \lineskip = \normallineskip
   \pretolerance=2000\tolerance=3000
   \hyphenpenalty=500\exhyphenpenalty=50
   \scriptfont0=\fiverm
   \hangindent 4.1mm\hangafter 1
   \noindent\hbox to 4.1mm{\hss$^{\the\fnnumber}$\kern 0.67mm}%
   \vrule width 0pt height 1\baselineskip depth 1.6pt \eightrm #2%
   \vphantom{p}}%
   \fi$^{\csname FnN#1\endcsname}\!$}
\line{\hfil OHSTPY-HEP-E-93-017}
\line{\hfil IKTP-DD/93-5}
\line{\hfil hep-ph/9312229}
\vskip 1cm

\title{The Semileptonic Decay Fraction of $B$-Mesons in the Light
of Interfering Amplitudes}

\authors {%
 K\:Honscheid\fn{grant}{Supported under
 DOE grant number DE-FG02-91-ER40690.}}%
\inst{Ohio State University, Columbus, Ohio, USA}

\authors {%
 K\:R\:Schubert, 
 R\:Waldi}
\inst{Institut f\"ur Kern- und Teilchenphysik\fn{DD}{%
Supported by the German Bundesministerium f\"ur Forschung
und Technologie, under contract number 056DD11P.},
Technische Universit\"at Dresden, Germany}

\vskip 6mm plus 1.5fil
}
{\ninefonts\rightskip 1cm\leftskip 1cm
\centerline{\bf Abstract}
\vskip 4pt plus 0.1pt minus 0.3pt
\noindent %
Consequences of the interference between spectator amplitudes
for the lifetimes and semileptonic decay fractions of $B^0$ and $B^+$
mesons are discussed.
Extracting these amplitudes from a fit
to 11 exclusive hadronic $B$ decay fractions
we find
$a_1 = 1.05 \pm 0.03 \pm 0.10$,
$a_2 =+0.227 \pm 0.012 \pm 0.022$,
an inclusive semileptonic decay fraction of
$(11.2\pm0.5\pm1.7)\%$, and a
lifetime ratio
$\tau(B^+) / \tau(B^0) = 0.83\pm0.01\pm0.01$.
\par\bigskip}

Although there has been significant progress in the
calculation of QCD corrections in the decays of heavy flavour mesons,
there are still some unsolved puzzles.
One of the most intriguing is the low semileptonic
decay fraction of $B$ mesons \cite{theory}.
We will show in this paper that the discrepancy between theory and
experiment
is considerably reduced by the
interpretation of recent CLEO
results \cite{cle} on hadronic decay fractions in the
framework of the spectator model with factorization.

The $D^0$-$D^+$ meson lifetime difference has been
satisfactorily reproduced in such a model by Bauer, Stech, and Wirbel
\cit{\ReF{bswhad},\ReF{bauer}}, using
interfering amplitudes of spectator diagrams
for $D^+$ hadronic two-body decays.
The same model predicts a negligible difference between the
decay rates for two body modes of $B^0$ and $B^+$ mesons.
However, in contrast to the $D$ system, only a small fraction of all
hadronic $B$ meson decays is included in the calculation so a conclusive
prediction of the lifetime ratio is not possible.
Using duality between the quark and the hadron pictures of strong
interactions, we will demonstrate how interfering spectator
amplitudes on the quark level may explain
the low
semileptonic decay fraction for $B$ decays, and at the same time predict
the lifetime ratio.

\midinsert
\figtop{2mm}
\figbody{\kern 50truept\sf
\hbox{%
\putl(0,118){\bf a)\kern 2em}%
\putl(0,-2){$d$}%
\putl(0,38){$\bbar$}%
\put(20,58){$W$}%
\putl(0,18){$\displaystyle{B^0\left\lbrace\vdistance{height 25truept
 depth 25truept}\right.}$\kern 0.5em}%
\vbox{\offinterlineskip
\hbox{\kern 15 truept\lower 20 truept\hbox{\feynlloop}\feynxline{40}}
\kern -20 truept
\hbox{\kern 35 truept\feynxline{40}}
\hbox{\feynxline{35}\feynvdash\feynxline{40}}
\kern 40 truept
\hbox{\feynxline{75}}
}
\put(0,-2){$d$}%
\put(0,38){$\cbar$}%
\put(0,78){$\dbar$}%
\put(0,118){$u$}%
\put(0,18){\kern 0.5em$\displaystyle{\left.\vdistance{height 25truept
 depth 25truept}\right\rbrace}$ hadron(s)}%
\put(0,98){\kern 0.5em$\displaystyle{\left.\vdistance{height 25truept
 depth 25truept}\right\rbrace}$ hadron(s)}%
}
\kern 130truept
\hbox{%
\putl(0,118){\bf b)\kern 2em}%
\putl(0,-2){$d$}%
\putl(0,38){$\bbar$}%
\putl(0,18){$\displaystyle{B^0\left\lbrace\vdistance{height 25truept
 depth 25truept}\right.}$\kern 0.5em}%
\put(20,58){$W$}%
\vbox{\offinterlineskip
\hbox{\kern 15 truept\lower 20 truept\hbox{\feynlloop}\feynxline{70}}
\kern -20 truept
\hbox{\kern 35 truept\feynxline{15}\kern 40 truept \feynxline{15}}
\hbox{\feynxline{35}\feynvdash\feynxline{15}\feyncross\feynxline{15}}
\kern 40 truept
\hbox{\feynxline{105}}
}
\put(0,-2){$d$}%
\put(0,38){$\dbar$}%
\put(0,78){$\cbar$}%
\put(0,118){$u$}%
\put(0,18){\kern 0.5em$\displaystyle{\left.\vdistance{height 25truept
 depth 25truept}\right\rbrace}$ hadron(s)}%
\put(0,98){\kern 0.5em$\displaystyle{\left.\vdistance{height 25truept
 depth 25truept}\right\rbrace}$ hadron(s)}%
}}
\figspace{3mm}
\figbody{\kern 50truept
\hbox{%
\putl(0,118){\bf c)\kern 2em}%
\putl(0,-2){$u$}%
\putl(0,38){$\bbar$}%
\put(20,58){$W$}%
\putl(0,18){$\displaystyle{B^+\left\lbrace\vdistance{height 25truept
 depth 25truept}\right.}$\kern 0.5em}%
\vbox{\offinterlineskip
\hbox{\kern 15 truept\lower 20 truept\hbox{\feynlloop}\feynxline{40}}
\kern -20 truept
\hbox{\kern 35 truept\feynxline{40}}
\hbox{\feynxline{35}\feynvdash\feynxline{40}}
\kern 40 truept
\hbox{\feynxline{75}}
}
\put(0,-2){$u$}%
\put(0,38){$\cbar$}%
\put(0,78){$\dbar$}%
\put(0,118){$u$}%
\put(0,18){\kern 0.5em$\displaystyle{\left.\vdistance{height 25truept
 depth 25truept}\right\rbrace}$ hadron(s)}%
\put(0,98){\kern 0.5em$\displaystyle{\left.\vdistance{height 25truept
 depth 25truept}\right\rbrace}$ hadron(s)}%
}
\kern 130truept
\hbox{%
\putl(0,118){\bf d)\kern 2em}%
\putl(0,-2){$u$}%
\putl(0,38){$\bbar$}%
\putl(0,18){$\displaystyle{B^+\left\lbrace\vdistance{height 25truept
 depth 25truept}\right.}$\kern 0.5em}%
\put(20,58){$W$}%
\vbox{\offinterlineskip
\hbox{\kern 15 truept\lower 20 truept\hbox{\feynlloop}\feynxline{70}}
\kern -20 truept
\hbox{\kern 35 truept\feynxline{15}\kern 40 truept \feynxline{15}}
\hbox{\feynxline{35}\feynvdash\feynxline{15}\feyncross\feynxline{15}}
\kern 40 truept
\hbox{\feynxline{105}}
}
\put(0,-2){$u$}%
\put(0,38){$\dbar$}%
\put(0,78){$\cbar$}%
\put(0,118){$u$}%
\put(0,18){\kern 0.5em$\displaystyle{\left.\vdistance{height 25truept
 depth 25truept}\right\rbrace}$ hadron(s)}%
\put(0,98){\kern 0.5em$\displaystyle{\left.\vdistance{height 25truept
 depth 25truept}\right\rbrace}$ hadron(s)}%
}}
\figspace{4mm}
\figcap{Spectator diagrams of hadronic
$B^0$ {\bf(a,b)} and $B^+$ {\bf(c,d)} decays. Diagrams {\bf(b,d)}
are often called colour suppressed since they lead to
stable hadrons if the colours of the combined quarks match
accidentally. Diagrams {\bf(a)} and {\bf(b)} for the $B^0$
decay lead to different final states, while
diagrams {\bf(c)} and {\bf(d)} for the $B^+$
decay can lead to the same hadronic final state and interfere.}
\FiG{diag}
\figbot
\endinsert

The spectator diagrams for hadronic $B^0$ and $B^+$ decay are
shown in Fig.~\FiGdiag.
Diagrams (a) and (b) are topologically
identical as are (c) and (d), but
the quarks are combined into final state hadrons
in a different way.
In $B^+$ meson decay, diagrams (c) and (d) can lead to the same
final state hadrons and hence the corresponding amplitudes
interfere.
The two diagrams for describing $B^0$ decays do not interfere, since
in (a) both final state hadrons are charged, while
in (b) both are neutral.
In the quantitative description,
two modifications of the weak decay amplitude lead to
diagrams (a) to (d):

\blist Owing to the
additional exchange of one or more gluons in parallel to
the $W$, effective neutral current contributions occur, with the same
$V-A$ structure as the charged current matrix element from pure $W$
exchange. These contributions would be the only source for diagrams
(b) and (d) in the
infinite colour limit $\Nc \to \infty$. They have been calculated
in next to leading log QCD approximation \cite{rueckl}, leading to
a coefficient $c_2 \approx -0.26$ at the $b$ mass scale
for 4 flavours and $\LQCD = 250\MeV$
\cite{neubert}.
Diagrams (a) and (c) are
enhanced by these QCD effects, leading to a coefficient
$c_1 \approx 1.12$ for this amplitude.

\blist Recombination of the mixed quark antiquark pairs
is possible if the colours match accidentally. This introduces
a factor $1/\Nc$ in the amplitude
relative to the diagrams where the
quarks are already in a colour singlet state,
leading to new coefficients
$$\eqalignno{
a_1 &= c_1 + {1\over \Nc} c_2 \approx 1.03 \cr
a_2 &= c_2 + {1\over \Nc} c_1 \approx 0.11 \cr
}$$
for diagrams (a,c) and (b,d), respectively.

Since
a fit to exclusive two body decay fractions of $D$ mesons
yields  the experimental results
$a_1 \approx c_1$ and $a_2 \approx c_2$
\cite{bswhad},
it has been argued that the $1/\Nc$ correction should be omitted,
corresponding to the limit $\Nc \to \infty$.
However, so far no convincing argument has been found to support this
proposition. It is also possible that $a_1$ and $a_2$ differ from the
QCD expectation because
$c_1$ and $c_2$ cannot be reliably calculated perturbatively.
The $b$ quark mass scale could be sufficiently
large for more reliable predictions.

\topinsert
\ninefonts
\tabtop{\hfil\?\hfil&
      |\hfil\?\hfil&
      |\hfil\?\hfil&
      |\hfil\?\hfil&}
\tabcap{Experimental averages and
theoretically predicted decay fractions
for hadronic $B$ decays,
assuming $|V_{cb}|^2\cdot \tau_{B} = 2.35\E{-15}\se$, and
$f_D = f_{D^*} = 220 \MeV$}
\tabspace
&&   decay&&
 exp. average $[\%]$&&
Neubert et al. \cite{neubert}&&
 Deandrea et al. \cite{dea}&
 \cr
\tabrule
&&$B^+ \rightarrow \Dbar^0 \pi^+$&&
$0.45 \pm 0.04 $&&
$0.265(a_1 + 1.230 a_2)^2$&& $0.268(a_1 + 1.16 a_2)^2$&
\cr
&&$B^+ \rightarrow \Dbar^0 \rho^+$&&
$1.10 \pm 0.18 $&&
$0.622(a_1 + 0.662 a_2)^2$&& $0.693(a_1 + 0.46 a_2)^2$&
\cr
&&$B^+ \rightarrow \Dbar^{*0} \pi^+$&&
$0.51 \pm 0.08 $&&
$0.255(a_1 + 1.292 a_2)^2$&& $0.268(a_1 + 1.71 a_2)^2$&
\cr
&&$B^+ \rightarrow \Dbar^{*0} \rho^+$&&
$1.32 \pm 0.31 $&&
$.70(a_1^2 + 1.49 a_1a_2 + .64 a_2^2)$&&
 $.92(a_1^2 + 1.31 a_1a_2 + .60 a_2^2 )$&
\cr
&&$B^+ \rightarrow \psi K^+$&&
$0.106 \pm 0.015$&&
$1.819\, a_2^2$&& $1.587\, a_2^2$&
\cr
&&$B^+ \rightarrow \psi K^{*+}$&&
$0.17 \pm 0.05$&&
$2.932\, a_2^2$&& $2.325\, a_2^2$&
\cr
\tabrule
&&$B^0 \rightarrow D^- \pi^+$&&
$0.26 \pm 0.04$&&
$0.264\, a_1^2$&& $0.268\, a_1^2$&
\cr
&&$B^0 \rightarrow D^- \rho^+$&&
$0.69 \pm 0.14$&&
$0.621\, a_1^2$&& $0.693\, a_1^2$&
\cr
&&$B^0 \rightarrow D^{*-} \pi^+$&&
$0.29 \pm 0.04$&&
$0.254\, a_1^2$&& $0.268\, a_1^2$&
\cr
&&$B^0 \rightarrow D^{*-} \rho^+$&&
$0.74 \pm 0.16$&&
$0.702\, a_1^2$&& $0.917\, a_1^2$&
\cr
&&$B^0 \rightarrow \psi K^0$&&
$0.069 \pm 0.022$&&
$1.817\, a_2^2$&& $1.587\, a_2^2$&
\cr
&&$B^0 \rightarrow \psi K^{*0}$&&
$0.146 \pm 0.029$&&
$2.927\, a_2^2$&& $2.325\, a_2^2$&
\cr
\tabbot
\TaB{brth}
\endinsert

The distinction between interfering amplitudes for the $B^+$
and non-interfering for the $B^0$ may only be valid for
two-body decays.
On the other hand,
many-body final states will most likely start as two colour
singlet quark antiquark pairs, including
intermediate massive resonances.
Interference between final states via different resonant
channels involves strong phases
which modify the rate for each individual final state
in a random way and disappear in the sum of all states.
It seems therefore reasonable to extend the model for exclusive
two body decays to
the majority of hadronic final states in an inclusive picture
at the quark level. We assume that the formation of two
colour singlets is the essential step of hadron production,
which is
taken into account quantitatively by $a_1$ and $a_2$.
We neglect modifications by
decays into baryon anti-baryon pairs, where our assumption is
not valid.

Experimentally, values for $a_1$ and $a_2$ have been obtained
from the measured partial rates of the $B$ meson decay modes
$D\pi , D\rho , D^* \pi ,
D^* \rho , J/\psi K$, and $ J/\psi K^*$.
Combining the experimental decay fractions
measured by the ARGUS \cite{arg}
and CLEO \cite{cle} experiments gives the averages listed in
Table~\TaBbrth.
We have used $\BR(D^0 \to K^-\pi^+) = (3.90\pm0.16)\%$
\cite{bur},
$\BR(D^+ \to K^-\pi^+\pi^+) = (9.1\pm1.4)\%$ \cite{mkiii88},
$\BR(D^{*+} \to D^0\pi^+) = (68.1\pm1.6)\%$ \cite{cleo92dd},
and
$\BR(\Jpsi \to l^+ l^-) = (5.9\pm0.2)\%$ \cite{mkiii92}.
The partial rates are determined under the assumption
of equal decay fractions of the $\Y4S$ into $B^+B^-$ and
$B^0\Bbar^0$ pairs, \<ie
$f^{+-}/f^{00} = 1$. This quantity is not well measured experimentally;
we assume in the following
$f^{+-}/f^{00} = 1.00 \pm 0.10$.

To estimate $a_1$ we
use $B^0$ decays into $D^-\pi^+$, $D^-\rho^+$, $D^{*-}\pi^+$
and $D^{*-}\rho^+$.
Using theoretical predictions from the model by
Neubert et al.\ \cite{neubert}
and
$|V_{cb}|^2\cdot \tau_{B} = 2.35\E{-15}\se$,
$f_D = f_{D^*} = 220 \MeV$
we obtain
$$
|a_1| = 1.03 \pm 0.05 \pm 0.10 \pm 0.05
$$
where the first error is statistical including uncertainties in
the $D^0$ and $D^+$ decay fractions, the second is from the
error on $V_{cb}\cdot \sqrt{\tau (B)}$, and the third comes from
the uncertainty on $f^{+-}/f^{00}$.
The model of Deandrea et al.\ \cite{dea} gives a similar answer,
$
|a_1| = 1.02 \pm 0.05 \pm 0.10 \pm 0.05
$.

$B \rightarrow J/\psi$ decays can be used to obtain an estimate
for $|a_2|$. Combining the experimental results in
Table~\TaBbrth\ with the
model of Neubert et al.\ and the
same factors as above, we find
$$
|a_2| =  0.23 \pm 0.011 \pm 0.02 \pm 0.01
$$
where the errors are given as for $|a_1|$ above.
The model of Deandrea et al.\ leads to
$
|a_2| = 0.25 \pm 0.013 \pm 0.02 \pm 0.01
$.
Within the errors, both models give the same answer. In the following
we use the model by Neubert et al.

\topinsert
\vbox{\advance\hsize -5truecm
\tabtop{\?\hfil&
      |\hfil\?\hfil&
      |\hfil\?\hfil&}
\tabcap{Experimental results and theoretical predictions for ratios of
$B^+$ and $B^0$ decay rates, scaled to $f_{D(D^*)}= 220\MeV$}
\tabspace
&&&& exp. average&& Neubert et al. \cite{neubert}&
\cr
\tabrule
&&$R_1=\frac{\Gamma(B^+ \rightarrow \Dbar^0\pi^+)}{\Gamma(B^0
\rightarrow D^+\pi^-)}$&&
$1.71 \pm 0.38$&&
$(1 + 1.23 a_2/a_1)^2 $&
\cr
\tabspace
&&$R_2=\frac{\Gamma(B^+ \rightarrow \Dbar^0\rho^+)}{\Gamma(B^0
\rightarrow D^+\rho^-)}$&&
$1.60 \pm 0.46$&&
$(1 + 0.66 a_2/a_1)^2 $&
\cr
\tabspace
&&$R_3=\frac{\Gamma(B^+ \rightarrow \Dbar^{*0}\pi^+)}{\Gamma(B^0
\rightarrow D^{*+}\pi^-)}$&&
$1.79 \pm 0.39$&&
$(1 + 1.29 a_2/a_1)^2 $&
\cr
\tabbot
\TaB{rat}
}\endinsert

This procedure using
decay modes that are only sensitive to
either the $a_1$ or the $a_2$ amplitude
does not reveal the relative sign between $a_1$ and $a_2$.
The sign can be obtained from $B^+ \rightarrow \Dbar^0$ and
$B^+ \rightarrow \Dbar^{*0}$
decays, which have contributions from both amplitudes.
A relative plus
sign between the $a_1$ and the $a_2$ amplitudes would cause
$\Gam(B^+ \rightarrow \Dbar^{(*)0} \pi(\rho)^+ )/
\Gam(B^0 \rightarrow D^{(*)-} \pi(\rho)^+) > 1$, while a minus sign
would correspond to ratios below 1.
The experimental results and
the model predictions for the decay ratios in the modes
$D \pi^-$, $D \rho^- $, and $ D^{*} \pi^-$
are given in Table~\TaBrat.
They show a clear preference for the positive sign.
The theoretical prediction for the decay
$B^+ \rightarrow D^{*0} \rho^+$ is too uncertain
\cite{rieckert}
to include this mode in the determination of
$a_1$ and $a_2$.
Taking ratios of $B^+$ and $B^0$
decays eliminates the uncertainties due to $|V_{cb}|$ but leaves
those originating from $\tau (B^+)/\tau (B^0)$ and
$f^{+-}/f^{00}$. The main difference between the models are details
of the $B\rightarrow \pi$ and $B \rightarrow \rho $ form factors.
The predictions also depend on the $D$ and $D^*$ meson decay
constants $f_D$ and $f_{D^*}$. Following Neubert et al.\ \cite{neubert}
we assume $f_D = f_{D^*} =
220\MeV$.
On the experimental side, the error due to the $D^0$ decay fractions
cancels in the ratios involving $B \rightarrow D^*$ decays.
A least squares fit with seven $D^{(*)}$ modes
from Table~\TaBbrth,
excluding only
$B^+ \rightarrow D^{*0} \rho^+$, gives
$$\eqalignno{
a_1 &= 1.04 \pm 0.05\,, \cr
a_2 &= 0.24 \pm 0.06\,, \cr
}$$
corresponding to the ratio
$$
a_2/a_1 = +0.23 \pm 0.06 \pm 0.05 \pm 0.05\,.
$$
The first error is the one standard deviation error from the fit, the
second error is from $f^{+-}/f^{00}$, and the third error comes
from the uncertainty in $\tau (B^+)/\tau (B^0) = 1.00 \pm 0.10$.

\topinsert
\figtop {3mm}
\startfigbody\kern 22truemm\sf
 \histogram{10.00truecm}{9.00truecm}{5000}{20000}{-1000}{5000}
{\histobarwidth 0.1pt\histocross(10000+10000-5000,0+5000-1000)}%
 \histoytick{    0}{1.4mm}{$\sf      0.0$}
 \histoytick{ 1000}{0.7mm}{}
 \histoytick{ 2000}{0.7mm}{}
 \histoytick{ 3000}{0.7mm}{}
 \histoytick{ 4000}{0.7mm}{}
 \histoytick{ 5000}{0.0mm}{$\sf      0.5$}
 \histoytick{ 2500}{0mm}{$\sf a_2$}
 \histoxtick{ 5000}{1.4mm}{$\sf      0.5$}
 \histoxtick{ 6000}{0.7mm}{}
 \histoxtick{ 7000}{0.7mm}{}
 \histoxtick{ 8000}{0.7mm}{}
 \histoxtick{ 9000}{0.7mm}{}
 \histoxtick{10000}{1.4mm}{$\sf      1.0$}
 \histoxtick{11000}{0.7mm}{}
 \histoxtick{12000}{0.7mm}{}
 \histoxtick{13000}{0.7mm}{}
 \histoxtick{14000}{0.7mm}{}
 \histoxtick{15000}{1.4mm}{$\sf      1.5$}
 \histoxtick{16000}{0.7mm}{}
 \histoxtick{17000}{0.7mm}{}
 \histoxtick{18000}{0.7mm}{}
 \histoxtick{19000}{0.7mm}{}
 \histoxtick{12499}{0mm}{\sf\lower 12pt\hbox{$
 \displaystyle\sf \tau_+/\tau_0$}}
{\plotdotspacing 1mm
 \histomove (11014,-1000)
 \histodraw (10830, -834)
 \histodraw (10400, -417)
 \histodraw (10000,    0)
 \histodraw ( 9629,  416)
 \histodraw ( 9287,  833)
 \histodraw ( 8973, 1250)
 \histodraw ( 8685, 1666)
 \histodraw ( 8422, 2083)
 \histodraw ( 8183, 2500)
 \histodraw ( 7965, 2916)
 \histodraw ( 7768, 3333)
 \histodraw ( 7590, 3750)
 \histodraw ( 7429, 4166)
 \histodraw ( 7285, 4583)
 \histodraw ( 7156, 5000)
}%
 \histomove (20000, 167)
 \histodraw (19799, 182)
 \histodraw (19453, 208)
 \histodraw (17034, 416)
 \histodraw (15110, 625)
 \histodraw (13546, 833)
 \histodraw (12249,1041)
 \histodraw (11157,1250)
 \histodraw (10226,1458)
 \histodraw ( 9423,1666)
 \histodraw ( 9062,1770)
 \histodraw ( 8724,1875)
 \histodraw ( 8408,1979)
 \histodraw ( 8110,2083)
 \histodraw ( 7831,2187)
 \histodraw ( 7567,2291)
 \histodraw ( 7319,2395)
 \histodraw ( 7084,2500)
 \histodraw ( 6861,2604)
 \histodraw ( 6651,2708)
 \histodraw ( 6451,2812)
 \histodraw ( 6261,2916)
 \histodraw ( 6080,3020)
 \histodraw ( 5908,3125)
 \histodraw ( 5744,3229)
 \histodraw ( 5588,3333)
 \histodraw ( 5438,3437)
 \histodraw ( 5295,3541)
 \histodraw ( 5159,3645)
 \histodraw ( 5028,3750)
 \histodraw ( 5000,3772)
 \histoput(20000,626){\lower 0.7ex\hbox{\ $\sf D^0\pi^+$}}%
 \histomove (20000,1086)
 \histodraw (19907,1093)
 \histodraw (19282,1145)
 \histodraw (18133,1250)
 \histodraw (16175,1458)
 \histodraw (14568,1666)
 \histodraw (13228,1875)
 \histodraw (12092,2083)
 \histodraw (11119,2291)
 \histodraw (10276,2500)
 \histodraw ( 9539,2708)
 \histodraw ( 9204,2812)
 \histodraw ( 8889,2916)
 \histodraw ( 8593,3020)
 \histodraw ( 8313,3125)
 \histodraw ( 8048,3229)
 \histodraw ( 7798,3333)
 \histodraw ( 7561,3437)
 \histodraw ( 7336,3541)
 \histodraw ( 7122,3645)
 \histodraw ( 6918,3750)
 \histodraw ( 6724,3854)
 \histodraw ( 6540,3958)
 \histodraw ( 6364,4062)
 \histodraw ( 6195,4166)
 \histodraw ( 6034,4270)
 \histodraw ( 5880,4375)
 \histodraw ( 5733,4479)
 \histodraw ( 5591,4583)
 \histodraw ( 5456,4687)
 \histodraw ( 5326,4791)
 \histodraw ( 5200,4895)
 \histodraw ( 5080,5000)
 \histomove (20000, 1916)
 \histodraw (16210, 2000)
 \histodraw (13254, 2083)
 \histodraw (11140, 2166)
 \histodraw ( 9556, 2250)
 \histodraw ( 8328, 2333)
 \histodraw ( 7812, 2375)
 \histodraw ( 7348, 2416)
 \histodraw ( 6930, 2458)
 \histodraw ( 6551, 2500)
 \histodraw ( 6206, 2541)
 \histodraw ( 5890, 2583)
 \histodraw ( 5601, 2624)
 \histodraw ( 5334, 2666)
 \histodraw ( 5089, 2708)
 \histodraw ( 5030, 2718)
 \histodraw ( 5001, 2723)
 \histodraw ( 5000, 2724)
 \histoput(20000,2103){\lower 0.7ex\hbox{\ $\sf \Jpsi K^+$}}%
 \histomove (20000, 2291)
 \histodraw (18267, 2333)
 \histodraw (15153, 2416)
 \histodraw (12880, 2500)
 \histodraw (11151, 2583)
 \histodraw ( 9793, 2666)
 \histodraw ( 8700, 2750)
 \histodraw ( 7802, 2833)
 \histodraw ( 7411, 2875)
 \histodraw ( 7053, 2916)
 \histodraw ( 6723, 2958)
 \histodraw ( 6418, 3000)
 \histodraw ( 6136, 3041)
 \histodraw ( 5875, 3083)
 \histodraw ( 5631, 3125)
 \histodraw ( 5405, 3166)
 \histodraw ( 5193, 3208)
 \histodraw ( 5092, 3229)
 \histodraw ( 5043, 3239)
 \histodraw ( 5018, 3244)
 \histodraw ( 5000, 3248)
 \histomove ( 5000, 1666)
 \histodraw (20000, 2500)
 \histoput(20000,2734){\lower 0.7ex\hbox{\ $\sf \Jpsi K^{*0}$}}%
 \histomove ( 5000, 2083)
 \histodraw (11271, 2500)
 \histodraw (18953, 2916)
 \histodraw (20000, 2968)
\endfigbody
\figspace{5mm}
\figcap{%
One standard deviation contours of $a_2$ versus $\tau(B^+) /\tau(B^0)$
at fixed $a_1 = 1.05$
for $B^+ \to \Jpsi K^+$,
$B^0 \to \Jpsi K^{*0}$,
and $B^+ \to \Dbar^0 \pi^+$ in the model of Neubert et al. \cite{neubert}.
Each $\Jpsi$ channel has a second band with negative $a_2$,
whereas
the $D\pi$ channel has only solutions with positive $a_2$.
The dotted line is the prediction from eq.~1.
}
\figbot
\FiG{plot}
\endinsert

The sign of $a_2/a_1$ turns
out to be positive in agreement with QCD and
$\Nc = 3$. The agreement between the $a_2$ values determined from
$J/\psi K^{(*)}$ and $D^{(*)} \pi (\rho)$ decays is not necessary
since the spectator quarks (graphs a to d in Figure 1) could have
a different influence in each decay mode. However, it supports
our basic assumption that all hadronic decays of $B$ mesons
via two-body decay modes are determined by the
same $a_1$ and $a_2$ coefficients.
Therefore, we use all
11 decay fractions in the fits described below.

Under our assumption of duality, the
coefficients $a_1$ and $a_2$ can be used to predict
hadronic and semileptonic
partial widths of the $B^+$ and $B^0$ mesons.
Since $b \to u$ and $b \to s$ transitions have been
shown to be very small,
we may safely neglect them in the following formulae;
$b \to u$ contributions
are included in the fits using $V_{ub}/V_{cb} = 0.08$.
In
Table~\nextTaB, we
consider only $b \to c$ spectator decays.
The phase space factors for $b \to c\,q_2 q_3$ are
$$\eqalignno{
I(r_c,r_2,r_3) &= 24\int_{(r_c+r_2)}^{(1-r_3)} (\xi^2-r_c^2-r_2^2)
(1+r_3^2-\xi^2) \cr
&\ \sqrt{[\xi^2-(r_c+r_2)^2][\xi^2-(r_c-r_2)^2]
[1-(r_3+\xi)^2][1-(r_3-\xi)^2]}
\ {d\xi \over \xi} \cr}
$$
with $r_i = m(q_i)/m(b)$.
In Table~\nextTaB, we give the relative factors
$PS = I(r_c,r_2,r_3)\?/I(r_c,0,0)$ which
have been calculated using
current masses
of $0.009$, $0.005$, $0.18$, $1.24$, and $4.65\GeV\byc^2$
for $d$, $u$, $s$, $c$, and $b$ quarks.
The perturbative
QCD correction for the semileptonic width is~\cite{cm78}
$$\Gam(\bbar \to \cbar e^+ \nu) =
\Gam_0 \cdot \left( 1 - {2\pi\over 3}\alphas + {25\over 6\pi}
\alphas\right) \approx
0.86\Gam_0
$$
for $\LQCD = 200\MeV$ and 4 flavours.

\topinsert
\tabtop{%
          $\?$\hfil&
  |$\?$\hfil&
  |\hfil$\?$\hfil&
  |\hfil$\?$\hfil&
  \tabvrule{1pt}$\?$\hfil&
  |\hfil$\?$\hfil&
  |\hfil$\?$\hfil&
  |\hfil$\?$\hfil&}
\tabcap{Contributions from all $b\to c$ spectator diagrams.
       Partial widths are obtained as $\Gam =
       \Gam_0(\bbar \to \cbar e^+\nu)\cdot
       \hbox{CKM}\cdot\hbox{QCD}\cdot PS$.}
\tabspace
&& B^+ (\bbar u ) &&$\hfil$\hbox{CKM}&& \hbox{QCD} && PS &&
  B^0(\bbar d ) && \hbox{CKM} && \hbox{QCD} && PS &\cr
&&\ \to &&           &&            &&           &&
 \ \to &&            &&            &&           &\cr
\tabrule
&& \cbar u\,e^+\nu &&        && 0.86  && 1.00 &&
  \cbar d\,e^+\nu &&         && 0.86  && 1.00 &\cr
&& \cbar u\,\mu^+\nu &&      && 0.86  && 0.99 &&
  \cbar d\,\mu^+\nu &&       && 0.86  && 0.99 &\cr
&& \cbar u\,\tau^+\nu &&     && 0.86  && 0.23  &&
  \cbar d\,\tau^+\nu &&      && 0.86  && 0.23  &\cr
&& \cbar u\ \dbar u &&|V_{ud}|^2=0.951&& 3(a_1+a_2)^2 &&
                                           1.00 &&
  \cbar d\ \dbar u &&|V_{ud}|^2      && 3a_1^2 &&1.00 &\cr
&&                 &&                &&        &&     &&
  \cbar u\ \dbar d &&|V_{ud}|^2      && 3a_2^2 &&1.00 &\cr
&& \cbar u\ \sbar u &&|V_{us}|^2=0.049&& 3(a_1+a_2)^2 &&
                                           0.98 &&
  \cbar d\ \sbar u &&|V_{us}|^2      && 3a_1^2 &&0.98 &\cr
&&                 &&                &&        &&     &&
  \cbar u\ \sbar d &&|V_{us}|^2      && 3a_2^2 &&0.98 &\cr
&& \cbar u\ \sbar c &&|V_{cs}|^2=0.949&& 3a_1^2 &&0.48 &&
  \cbar d\ \sbar c &&|V_{cs}|^2      && 3a_1^2 &&0.48 &\cr
&& \cbar c\ \sbar u &&|V_{cs}|^2     && 3a_2^2 &&0.48 &&
  \cbar c\ \sbar d &&|V_{cs}|^2      && 3a_2^2 &&0.48 &\cr
&& \cbar u\ \dbar c &&|V_{cd}|^2=0.049&& 3a_1^2 &&0.49 &&
  \cbar d\ \dbar c &&|V_{cd}|^2      && 3a_1^2 &&0.49 &\cr
&& \cbar c\ \dbar u &&|V_{cd}|^2     && 3a_2^2 &&0.49 &&
  \cbar c\ \dbar d &&|V_{cd}|^2      && 3a_2^2 &&0.49 &\cr
\tabspace
\tabbot
\TaB{corr}
\endinsert

 From the factors in Table~\TaBcorr\ we obtain
the following total widths, normalized
to the lowest order semileptonic width
$\Gam_0(b \to c e^- \anu)$
$$\eqalignno{
{\Gam(B^+) /\Gam_0} &=
 1.91 + 4.44 (a_1^2 + a_2^2) + 5.99 a_1 a_2\,, \cr
{\Gam(B^0) /\Gam_0} &=
 1.91 + 4.44 (a_1^2 + a_2^2)\,. \cr
}$$
Using these widths, we can calculate two important quantities.
\blist The average
semileptonic decay fraction of $B^0$ and $B^+$,
$$
\BR(B \to e \nu X) =
{1 \over
 2.22 + 5.16 (a_1^2 + a_2^2) + 3.49 a_1 a_2}\,,
$$
decreases if $a_2$ changes sign from negative to positive.
\blist The lifetime ratio
$$
\tau(B^+) / \tau(B^0) =
 1 - {a_1 a_2 \over
 0.32 + a_1 a_2 + 0.74 (a_1^2 + a_2^2)}
\eqno(\rm\theeqnum\QnR{tt})
$$
is larger than 1 for negative and
smaller than 1 for positive values of $a_2$.
The relation between the lifetime ratio and
$a_2$ is shown in Fig.~\FiGplot\ as dotted line for fixed
$a_1 = 1.05$.

To give consistent results, we determine $a_1$ and $a_2$
in a fit to the 11 decay fractions used above,
replacing the assumption of equal $B^+$ and $B^0$ lifetimes
with the inclusive prediction in
eq.~\QnRtt\ to rescale the theoretical expectations
for $B^+$ and $B^0$ decays individually.
This fit gives $\chi^2 = 11.6$ for 8 degrees of freedom, and
$$\eqalignno{
a_1 &= 1.05 \pm 0.03 \pm 0.10\cr
a_2 &= 0.227 \pm 0.012 \pm 0.022\cr
}$$
which implies
$$\eqalignno{
\BR(B \to e \nu X) &= (11.2\pm0.5\pm1.7)\% \cr
\tau(B^+) / \tau(B^0) &= 0.83\pm0.01\pm0.01\,,\cr
}$$
where the first error is statistical including uncertainties in
the $D^0$ and $D^+$ decay fractions, and
the second is from the
error on $V_{cb}\cdot \sqrt{\tau (B)}$. The uncertainty
on $f^{+-}/f^{00}$ yields a negligible error.
The semileptonic decay fraction is further reduced if we assume
a small contribution of penguin decays.
Assuming this fraction to be $2.5\%$ leads to $\chi^2 = 11.3$ and
$\BR(B \to e \nu X) = 10.9\%$, while all errors and the values of
$a_1$, $a_2$ and
$\tau(B^+) / \tau(B^0)$ remain essentially unchanged.

Our fit results agree
reasonably well with present experimental values:
The average $B^+$ and $B^0$ semileptonic decay fraction
is
$\BR(B \to l \nu X) = (10.2\pm0.3)\%$. This is the average of all
inclusive $e$ and $\mu$ results \cite{slbr}
from data taken
on the $\Y4S$, where only these two
types of $b$-flavoured mesons are produced.
The result for all $b$-hadrons obtained at LEP is
only slightly higher,
$\BR(b \to l \nu X) = (11.0\pm0.5)\%$ \cite{dani}.
The lifetime ratio from
LEP and CDF is \cite{dani}
$
\tau(B^+) / \tau(B^0) = 1.07\pm0.12
$.

Assuming duality and constructive interference between spectator
amplitudes we are able to explain the low experimental value
for the semileptonic decay fraction of $B$ mesons.
The experimental data on the lifetime ratio
are not yet sufficiently precise to either confirm
or falsify our prediction that the $B^+$ has a shorter mean
life than the $B^0$.
However, a small contribution from annihilation
diagrams, which enhance $B^0$ decays, could
raise the expected value.
This would also bring the
semileptonic decay fraction even closer to the experimental average.

{\bf Acknowledgements}. We thank
V\:Rieckert and B\:Stech for helpful
discussions on hadronic decays, and
T\:E\:Browder for useful comments.

{\frenchspacing
\hdrni{References}

\bibitem{theory}%
G\:Altarelli and S\:Petrarca, \PL B261, 303 (1991);
\newline
I\:I\:Bigi et al., \PL B293,
430 (1992) and erratum \CiT{ibid.} B297, 477 (1993);
\newline
I\:I\:Bigi, B\:Blok, M\:A\:Shifman, A\:Weinshtein,
CERN-TH 7082/93 (1993).

\bibitem{cle}D\:Bortoletto et al. (CLEO), \PR D45, 21 (1992);\newline
M\:S\:Alam et al. (CLEO),
CLEO-CONF 93-33, contributed paper to the XVIth Lepton Photon
Symposium, Ithaca 1993.

\bibitem{bswhad}M\:Bauer, B\:Stech, M\:Wirbel,
\ZP C34,103(1987).

\bibitem{bauer}M\:Bauer, Dissertation,
Heidelberg 1987 (unpublished).

\bibitem{rueckl}R\:R\"uckl, Habilitationsschrift, CERN-Print 83-1063
(1983).

\bibitem{neubert}M\:Neubert, V\:Rieckert, B\:Stech in `Heavy Flavors',
ed.\ by A\:J\:Buras and M\:Lindner, p.~286,
World Scientific Publ.\ 1992.

\bibitem{dea}A\:Deandrea et al., Preprint UGVA-DPT 1993/07-824.

\bibitem{arg}H\:Albrecht et al. (ARGUS), \ZP C48,543 (1991).

\bibitem{bur}P\:R\:Burchat,
Proceedings of the 5th Int.\ Symposium
on Heavy Flavour Physics, Montreal, Canada, July 1993.

\bibitem{mkiii88}J\:Adler et al. (Mark III), \PRL 60,89(1988).

\bibitem{cleo92dd}F\:Butler et al. (CLEO), \PRL 69,2041 (1992).

\bibitem{mkiii92}D\:Coffman et al. (Mark III), \PRL 68, 282 (1992).

\bibitem{rieckert}V\:Rieckert, priv. communication.

\bibitem{cm78}N\:Cabibbo, L\:Maiani, \PL B79,109 (1978).

\bibitem{slbr}K\:Wachs et al. (Crystal Ball),
\ZP C42,33 (1989);\newline                               
H\:Albrecht et al. (ARGUS), \PL B249, 359 (1990);\newline 
C\:Yanagisawa et al. (CUSB), \PRL 66, 2436 (1991);\newline
S\:Henderson et al. (CLEO), \PR D45, 2212 (1992);\newline 
H\:Albrecht et al. (ARGUS),                              
\PL B318, 397 (1993).

\bibitem{dani}M\:Danilov, Proceedings of the European Physical
Society
Conference, Marseille, France, July 1993.

}
\bye